# SELF-STIMULATED UNDULATOR KLYSTRON

E.G Bessonov, A.L. Osipov, Lebedev Phys. Inst. RAS, Moscow, Russia
A.A.Mikhailichenko, Cornell University, CLASSE, Ithaca, NY 14853, U.S.A.

*Abstract*
The Self Stimulated Undulator Klystron (SSUK) and its possible applications in the Particle Accelerator Physics, incoherent Self-Stimulated Undulator Radiation Sources (SSUR) and Free-Electron Lasers (FEL) are discussed.

## INTRODUCTION

The system of two undulators located one by one in a sequence at some distance along the straight line is called an Undulator Klystron (UK). It was invented by R.M.Phillips in 1960 for generation of spontaneous coherent UR [1]. In the first undulator (modulator) the electron beam was modulated by energy in the field of copropagated wave, then in a straight section it was grouped in the bunches and in the second undulator (radiator) the bunched beam emitted a coherent undulator radiation (UR) on the lowest or higher harmonics. Inverse FEL-accelerator scheme and tapered undulators were suggested there as well. Note that the both spontaneous incoherent and coherent UR sources were suggested by V.L.Ginsburg in 1947 [2]. Later the spontaneous coherent UR sources were named by parametric (superradiant, pre-bunched) FELs [3], [4]. Below we will use more suitable term "pre-bunched" suggested by A.Gover [5]. The UR emitted in the UK consisted of $N_u$ undulators located at some distances along a straight line was investigated in [6], [7]. The UK with a dispersion element located in its straight section for enhancement the bunching process for the ultrarelativistic particles was called an Optical Klystron (OK); it was suggested in 1977 [8].

SSUK is further modification of UK with controlled delay of the Undulator Radiation Wavelets (URWs) moving between the undulators [9], [10]. The optical delay line is arranged with the mirrors and lenses. It serves for proper phasing of the URWs with particles for their further interaction in the following undulator. The special magnet system installed between the undulators (kicker) serves for separation of the URWs from the particle beam, see Fig.1. The URWs and particle beams are focused back to the location at the entrance of the downstream undulator. The optical and particle's delays are chosen so that the particle enters the following undulator in a decelerating phase at the front edge of its own URW, emitted in a preceding undulator. Under such conditions the superposition of the URW emitted in the first undulator and the URW emitted in the following one occurs, which yields the field growth $\sim N_u$ and the energy density growth in emitted radiation becomes $\sim N_u^2$. So the Self-Stimulated UR (SSUR) is emitted by each particle in the SSUK in the self-fields of its own wavelets emitted at earlier times in the upstream undulators.

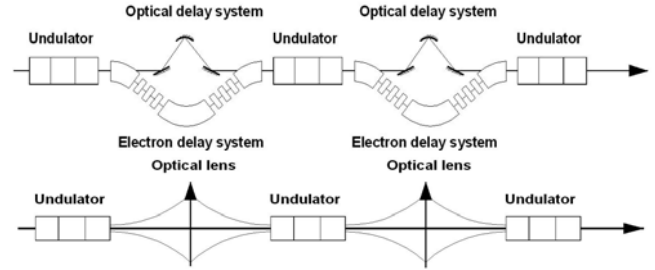

Figure 1. Scheme of the installation and its equivalent optical analog.

We would like to underline here, that in OK there is no requirements that the particle should interact with its own wavelet, so here is the principal difference between OK and SSUK. The SSUK scheme is similar to ones considered in [11], [12]. It has two undulators also but does not include an X-ray monochromator or an amplifier located in between. The first undulator stage in [11] operates as the FEL amplifier in the SASE linear regime. After exiting out from the first undulator the electron bunch is guided through a non-isochronous bypass and the X-ray beam enters the monochromator. The main function of this bypass is to suppress the modulation of the electron beam induced in the first undulator. At the entrance in the second undulator the radiation power from the monochromator dominates significantly over the shot noise and the power determined by the residual electron bunching. As a result, the second stage of the FEL amplifier operates in a steady-state regime. Phase correlations of the URWs emitted in the first and second undulators by the same particle was not considered in [11]. The URW emitted in the first undulator in [12] was amplified and used in the second undulator for production a coherent kick to the particle for cooling of the beam. URW emitted in the second undulator was not used.

Using the SSUK in our case means the generation of coherent URWs by every particle of the beam in the system of $N_u$ undulators of the SSUK under conditions of incoherent emission of coherent URWs by particles of the beam (incoherent superposition of coherent URWs emitted by $N$ particles of the beam in $N_u$ undulators of the SSUK). These elements are playing a key role in our proposal of SSUK. Usage of SSUK in FEL where the coherence among the particles of the beam exists already is useful as well.

## MAGNETIC LATTICE PROPERTIES OF THE SSUK

We considered the case when the optical delays are tuned so that the wavelets emitted by the particles are congruent and all particles stay at decelerating phase. For this pur-poses the beam delay system in the SSUK must be quasi-isochronous. To be optimally effective, the optical part of a system must use appropriate focusing elements such as lenses and/or focusing mirrors. The mirrors and lenses form a crossover in the middle of the undulators with the Rayleigh length $Z_R \cong M\lambda_u/2$, where $\lambda_u$ is the undulator period, $M$ is the number of undulator periods.

The URWs emitted by each electron on the harmonic with the number $m$ are overlapped effectively at the exit of the SSUK by the superposition one by another if theirs longitudinal shifts satisfy the following condition

$$\Delta l = |c \cdot \Delta T_{e,URW} - n\lambda_m| \ll \lambda_m/2, \qquad (1)$$

where $\lambda_m = \lambda_1/m$ is the wavelength of the UR emitted by the electron on the $m$-th harmonic in the direction of its average velocity, $\Delta T_{e,URW} = T_e - T_{URW}$ is the difference between the entrance time of the URW and the electron to the next undulator, $T_{URW} = (L_{uu} + \Delta l)/c = const$, $T_e = T_e(\varepsilon, K_{kick}, \theta_{in})$, $L_{uu}$ is a distance between the undulators, $L_{uu} + \Delta l$ is the length of the light way in the optical delay system, $\varepsilon$ is the particle energy, $K_{kick}$ is the deflection parameter of the kicker (the kicker can be considered as a half - period undulator), $\theta_{in}$ is the initial angle between the particle velocity and the UK axis, $n = 0, \pm 1, \pm 2,... |n| \leq mM$ is the synchronicity condition number.

The electron pass on the way between undulators for the time $T_e = \tilde{L}_{uu}/v$, where $\tilde{L}_{uu} = \int_0^{L_{uu}} ds/\cos[\theta_{in} + \Delta\theta(s)]$ is the length of particle's trajectory between the undulators, $v$ is the particle velocity, $\Delta\theta(s) = \int_0^s [eB(s)/\beta\varepsilon]ds$ is the bending angle along the kicker's axis at the position $s$, $\beta = v/c$, $B(s)$ is the kicker's transverse bending magnet field strength. In the relativistic case $\beta = v/c \simeq 1 - 1/2\gamma^2$, $|\theta_{in} + \Delta\theta(s)| \ll 1$, $\cos[\theta_{in} + \Delta\theta(s)] \simeq 1 - (\theta_{in} + \Delta\theta(s))^2/2$ the values

$$\tilde{L}_{uu} = L_{uu}(1 + \frac{\theta_{in}^2}{2}) + \frac{e^2}{2m_e^2c^4\beta^2\gamma^2}\int_0^{L_{uu}}[\int_0^s B(s)ds]^2 ds,$$

$$T_e = \tilde{L}_{uu}/v = \frac{L_{uu}}{c}(1 + \frac{\theta_{in}^2}{2} + \frac{1}{2\gamma^2}) + \frac{e^2\int_0^{L_{uu}}[\int_0^s B(s)ds]^2 ds}{2m_e^2c^5\beta^3\gamma^2},$$

$$\delta T_e = T_e - T_{e,0} = \frac{L_{uu}}{2c\gamma^2}(1 + K_{kick}^2 + \vartheta_{in}^2), \qquad (2)$$

where $\gamma = \varepsilon/m_ec^2 \gg 1$ is the particle's relativistic factor, $K_{kick} = e\sqrt{\int_0^{L_{uu}}[\int_0^s B(s)ds]^2 ds/L_{uu}}/m_ec^2$, $\vartheta = \theta\gamma$, $T_{e,0} = T_e[\theta_{in} = 0, B(s) = 0]$. The shifts of URWs, according to (1), (2) are $\Delta l = c(\partial \delta T_e/\partial \gamma)\Delta\gamma_b + c(\partial(\delta T_e)/\partial \vartheta_{in})\sigma_b' = -2c(\delta T_e)(\Delta\gamma_b/\gamma) + cT_e(\sigma_b')^2/(1 + K_{kick}^2) \ll \lambda_m/2$, where $\Delta\gamma_b$ and $\sigma_b'$ are the energy and angular spreads of the particle beam. It follows from here that the requirements to the beam parameters should be

$$\frac{\Delta\gamma_b}{\gamma} \ll \frac{\lambda_m\gamma^2}{2L_{uu}(1 + K_{kick}^2)} = \frac{\lambda_u(1 + K_u^2)}{4mL_{uu}(1 + K_{kick}^2)},$$

$$\sigma_b' \ll \sqrt{\frac{\lambda_m}{L_{uu}}} = \frac{1}{\gamma}\sqrt{\frac{\lambda_u(1 + K_u^2)}{2mL_{uu}}}, \qquad (3)$$

where $K_u$ is the undulator deflection parameter. We took into account the equation for the emitted UR wavelength $\lambda_m = \lambda_u(1 + K_u^2)/2m\gamma^2$. In this case the local slip factor $\eta_{c,loc} = (\gamma/T_e)\partial(\delta T_e)/\partial\gamma = (1 + K_{kick}^2)/\gamma^2$.

The requirements to the beam parameters for SSUK (3) are much easier than the ones for SSUR source based on the storage ring [10]. Note that a small slip factor system of two undulators separated by the bending magnetic system was used to study the radiation coherency conditions in optical region [13], [14]. It means that technical realization of tuning of the URWs is possible in the optical and even harder wavelength regions.

For a kicker with small bending magnet lengths ($l_{b1} = l_{b3} = l_{b2}/2 \ll L_{uu}$), the value $K_{kick} = \vartheta_b$, where $\theta_b = eBl_{b1}/mc^2\gamma \simeq 6 \cdot 10^{-4} B[gs] \cdot l_{b1}[cm]/\gamma$ is the bending angle of the first and third kicker magnets (we neglected the influence of magnetic fields of the quadrupole lenses). In this case the orbit will be deviated from the SSUK axis at its center by the value $a = L_{uu}\theta_b/2$. It must be 5-10 times higher than the rms particle beam size $\sigma_b$.

*Example.* Let the SSUK has 2 identical helical undulators with the period $\lambda_u = 3$ cm, number of periods $M = 30$, the distance between the undulators $L_{uu} = 1$ m, bending magnet lengths $l_{b1} = l_{b3} = l_{b2}/2 = 3$ cm, $\gamma = 10^3$, the deflection parameters $K_u = 2$, the transverse beam dimension $\sigma_b \simeq 1$ mm.

In this example the bending angle of the first and third kicker magnets $\theta_b = 2 \cdot 10^{-2}$, the deviation of orbit in the center of the SSUK $a = \theta_b L_{uu}/2 = 1$ cm, $K_{kick} = 20$, the bending magnet field strength $B \simeq 1.1 \cdot 10^4$ gs, the local slip factor $\eta_{c,loc} = 4 \cdot 10^{-4}$. The electron beam for every $n$ and $m=1$ must have the energy spread $\Delta\gamma_b/\gamma < 10^{-4}$, the angular spread $\sigma_b' < 2.7 \cdot 10^{-4}$.

## POSSIBLE APPLICATIONS
*Spontaneous incoherent UR sources based on SSUK*

All properties of spontaneous incoherent radiation emitted by the particle beam in an undulator and in the SSUK based on such undulators under main synchronicity condition $n = 0$ are identical, except intensity, which becomes higher by $N_u^2$ times. If the centers of URWs at the exit of the last SSUK undulator are displaced in the transverse direction inside some area with a dimension $d > \lambda_m \gamma$ then the additional degree of directionality will appear in the UR beam: $\Delta\theta \sim \lambda_m / d < \sqrt{1+K_u^2}/\gamma$. At the same time an increase in the power will be lesser than $N_u^2$ (phased antenna array analogy in prebunched FEL). It follows from the general theory of such FELs [15]-[16]. Two bending magnets with opposite polarity located between the undulators can be used for the transverse displacements of URWs. If URWs is emitted by an electron at the collateral synchronicity conditions $-mM < n \le mM$ are shifted in the longitudinal direction at the exit of the last undulator by the distances $\pm\lambda_m, \pm 2\lambda_m ... \pm mN_u\lambda_m$ then the additional directionality $\Delta\theta \simeq \sqrt{1+K_u^2}/\gamma \sqrt{min\{M, N_u\}}$ appears in the UR emitted by every electron (director-type antenna analogy) [15], [16]. Moreover the intensity will be increased by $min\{N_u^2, M^2\}$ times for $n \ll mM$. The intensity will be dropped and the monochromaticity will be increased $N_u$ times if the number $n \to mM$ ($n \le mM$). The angular spread of the beam in this case must be small $\sigma_b' < \Delta\theta$.

In both cases we deal with the self-stimulated UR for every particle of the beam and spontaneous incoherent UR between particles of the beam. There is no requirement for the coherence in radiation among the different electrons in the bunch like it is required for the prebunched FELs. The stimulated process of radiation for each electron is going in the undulator with their own URW fields only.

The considered phenomena of the power, directionality and monochromaticity increase in the SSUK installed in a storage ring or in the linear accelerators and recirculators takes place both for broad band and narrow band mirrors used in the SSUK in the optical up to to X-ray regions.

The accuracy of the SSUK lattice tuning is

$$\Delta l \ll \frac{\lambda_u (1+K_u^2)}{mN_u(1+K_{kick}^2)}. \qquad (4)$$

It follows from the necessity to maintain by the optical delay line the distance between next URW relative to the previous one with the accuracy of $\Delta l' \ll \lambda_m / N_u$.

## Free-electron lasers

*1. Prebunched FELs.* In spontaneous incoherent SSUR sources the URWs are emitted by each particle independently from the other particles of the beam. Single micro bunch with the number of particles $N_1$ and the length $l_{mkb} \ll \lambda_m$ is equivalent to one particle with the charge $eN_1$. The trains of such micro bunches in the prebunched SSUK FEL regime can be used here. In this case the power of the emitted coherent radiation is $P^{coh} \approx P^{incoh} N_1^2 N_u^2$, where $P^{incoh}$ is the power of incoherent radiation of the unbunched beam emitted in one undulator.

The system of a modulator undulator (in a combination with the driving laser beam) and the radiator SSUK installed in a storage ring (as well as in ordinary or energy recovering linacs and recirculators) tuned on the main or higher harmonics of the microbunched beam can be used by analogy with the scheme considered in [17].

*2. Ordinary FELs.* Using SSUK at the condition (1) in ordinary FELs will permit to decrease the threshold current of FELs and to increase their power.

## Cooling of particle beams

*1. Optical cooling.* Usage of SSUK klystrons as pickups for optical stochastic cooling [18], [12] and enhanced optical cooling of particle beams [19] will permit to increase the number of photons in the sample $N_u^2$ times and the cooling rate $N_u$ times. Using SSUK both as pickup and kicker undulator will increase the rate of cooling $N_u^2$ times.

*2. Cooling based on incoherent SSUR.* If the revolution period of a particle in a storage ring is multiple to the round trip period of the URWs emitted by the particle in the undulator installed in the straight section of the ring (see Fig. 2) and circulating in the optical resonator then such URWs will be effectively stored and overlapped in the resonator under resonance conditions in some energy interval [10]. Frictional losses of energy and cooling appear in this interval. Using an optical amplifier in the optical resonator system, switching it on for a short time (dozens of particle revolutions) and switching it off for a small number of revolutions can lead to the high rate of the particle energy loss, its strong dependence versus particle energy and, following the analogy with the Robinson theorem [21], to the high rate of particle beam cooling in the longitudinal phase space. In this case the losses of the energy in the undulator occur with equal probability and independently on the sign of transverse deviation of the particle from its instantaneous orbit. The jumps of the particle amplitudes of betatron oscillations have different signs and hence in the first approximation the cooling in the transverse plane is absent. Using SSUK instead of an undulator may increase the cooling rate of particles in the storage ring. We suppose that the current of the particle beam is less than a threshold current for such FEL-like scheme. Cooling is possible in such a way at the main and collateral synchronicity conditions as well [10].

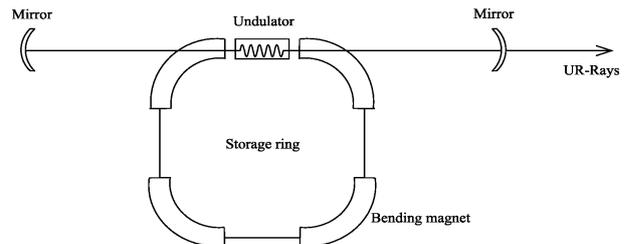

Figure 2. Schematic diagram of SSUR installation built around a storage ring.

The energy interval of the main and collateral synchronicity conditions depends on the slip factor of the ring. Quasi-isochronous storage rings are desirable in this case.

**CONCLUSION**

We hope that the SSUK can be used as a high efficiency pickup for cooling of the proton, muon and ion beams in the storage rings, as highly effective SSUR source based on ordinary and compact quasi-isochronous storage rings, ordinary and Bragg resonators capable generation both in the short and continuous, quasi-monochromatic light beams in the optical to X-ray regions. It can be used effectively both in the ordinary and prebunched FELs as well [10], [20]. Using SSUK in proposed energy recovery linacs, International linear collider and FEL sources will permit to enhance novel X-ray quantum optics experiments in the femtosecond regime and generate $\gamma$ rays carring the Orbital Angular Momentum for nuclear and particle physics [22].

This work was supported in part by RFBR under Grants No 09-02-00638a, 09-02-01190a.

**REFERENCES**

[1] R.M.Phillips. "The Ubitron, a High-Power Traveling Wave Tube Based on Periodic Beam Interaction in Unloaded Waveguide", IRE Transactions on Electron Devices, 1960, vol. ED-7, No 4, p.231-237.

[2] V.L.Ginsburg, "On Emission Micro Radio Waves and their Absorption in Air", Izvestia Academii Nauk SSSR, ser. Physicheskaia, 1947, v.11, No 2, p. 165.

[3] E.G.Bessonov, "Theory of Parametric Free - Electron Lasers", Sov. Journal Quantum Electron. 1986, v.16, N8, p.1056.

[4] E.G.Bessonov, "Parametric Free - Electron Lasers", Nucl Instr. Meth., 1989, A282, p.442.

[5] M. Arbel, A. Abramovich, A. L. Eichenbaum, A. Gover, H. Kleinman, Y. Pinhasi, and I. M. Yakover. "Superradiant and Stimulated Superradiant Emission in a Prebunched Beam Free-Electron Maser", PRL V.86, No 12, 2001, p. 2561-2564.

[6] E.G.Bessonov. "Undulators, Undulator Radiation, Free-Electron Lasers", Proc. Lebedev Phys. Inst., Ser.214, 1993, p.3-119, Chief ed. N.G.Basov, Editor-in-chief P.A.Cherenkov.

[7] E.G.Bessonov, "Peculiarities of Harmonic Generation in a System of Identical Undulators", Nucl. Instr. Meth. A 341 (1994), ABS 87; http://www.sciencedirect.com/science?_ob=MImg&_imagekey=B6TJM-470F3WY-J0-1&_cdi=5314&_user=492137&_pii=0168900294904596&_orig=search&_coverDate=03%2F01%2F1994&_sk=996589998&view=c&wchp=dGLbVzz-zSkzV&md5=8309b9f8f17e8b2ad6263367db281526&ie=/sdarticle.pdf .

[8] N.A.Vinokurov, A.N.Skrinsky. Preprint INP N°77-59, Novosibirsk (1977).

[9] E.G.Bessonov, M.V.Gorbunkov, A.A.Mikhailichenko, A.L.Osipov, "Self-Stimulated Emission of Undulator Radiation", Journal of Instrumentation JINST_012P_0510, 2010, p.1-4; arXive: http://arxiv.org/abs/1003.3747.

[10] E.G.Bessonov, M.V.Gorbunkov, A. Mikhailichenko, A.L.Osipov, A.V.Vinogradov, "Self-Stimulated Undulator Radiation and its Possible Applications", http://arxiv.org/ftp/arxiv/papers/1009/1009.3724.pdf.

[11] J.Feldhaus, E.L.Saldin, J.R.Schneider, E.A. Schneidmiller, M.V. Yurkov, "Possible Application of X-ray Optical Elements for Reducing the Spectral Bandwidth of an X-ray SASE FEL", Nuclear Instruments and Methods in Physics Research A 393 (1997) 162-166.

[12] M.S.Zolotorev, A.A.Zholents, "Transit-Time Method of Optical Stochastic Cooling", Phyas. Rev E, V. 50, No 4, 1994, p. 3087-3091.

[13] G.N.Kulipanov, V.N.Litvinenko, A.S.Sokolov, N.A.Vinokurov, "On Mutual Coherency of Spontaneous Radiation from two Undulators Separated by Achromatic Bend", IEEE Journal of Quantum Electronics, VOL. 27, No 12, 1991, p. 2566-2568.

[14] N.G.Gavrilov, G.N.Kulipanov, V.N.Litvinenko, I.V.Pinaev, V.M.Popik, I.G.Silvestrov, A.N.Skrinsky, A.S.Sokolov, N.A.Vinokurov, P.D.Vobly, "Observation of Mutual Coherency of Spontaneous Radiation from two Undulators Separated by Achromatic Bend", IEEE Journal of Quantum Electronics, VOL. 27, No 12, 1991, p. 2569-2571.

[15] D.F.Alferov, Yu.A.Bashmakov, E.G.Bessonov, "Theory of Undulator Radiation I", Sov. Phys. Tech. Phys. 1978, v.23, N8, p.902-904.

[16] D.F.Alferov, Yu.A.Bashmakov, E.G.Bessonov, "Theory of Undulator Radiation II", Sov. Phys. Tech. Phys. 1978, v.23, N8, p.905-909. Preprint FIAN No 163, 1976.

[17] E.G.Bessonov, "A Method of Harmonic Generation in a Storage Ring based FEL", Proc. of 21st Internat. Free Electron Lasers Conf., Aug.23-28, 1999, Hamburg, Germany, p.II-51 - II-52.

[18] A.A. Mikhailichenko, M.S. Zolotorev, "Optical Stochastic Cooling", Phys. Rev. Lett.71: 4146-4149, 1993.

[19] E.G. Bessonov, M.V. Gorbunkov, A.A. Mikhailichenko, "Enhanced Optical Cooling System Test in a Muon Storage Ring", Phys. Rev. ST Accel. Beams 11, 011302 (2008).

[20] E.G.Bessonov, M.V.Gorbunkov, A.A.Mikhailichenko, A.L.Osipov, "Self-Stimulated Undulator Radiation Sourses", XXII Russian Particle Accelerator Conference RuPAC-2010, September 27 – October 1, 2010, Protvino, Moscow Region, Russia, WEPSB004, p.181-183, http://accelconf.web.cern.ch/AccelConf/r10/papers/wepsb004.pdf.

[21] E.G.Bessonov. " The Evolution of the Phase Space Density of Particle Beams in External Fields", Proceedings of the Workshop On Beam Cooling and Related Topics, COOL 2009, Lanzhou, China, p. 91-93, 2009.
http://cool09.impcas.ac.cn/JACoW/papers/tua2mcio02.pdf;
http://lanl.arxiv.org/abs/0808.2342;
http://arxiv.org/ftp/arxiv/papers/0808/0808.2342.pdf.

[22] S.Sasaki and I. McNulty," Proposal for Generating Brilliant X-Ray Beams Carrying Orbital Angular momentum", Phys. Rev. Lett., V.100, 124801 (2008).